# Bridging Classical Molecular Dynamics and Quantum Foundations for Comprehensive Protein Structural Analysis

Don ROOSAN [a, 1], Rubayat KHAN [b], Tiffany KHOU [c], Saif NIRZHOR [d], Fahmida HAI [e] and Brian PROVENCHER [f]

[a] *Department of Computer Science, Merrimack College, 315 Turnpike St, North Andover, MA 01845, United States*
[b] *University of Nebraska Medical Center, S 42nd &, Emile St, Omaha, NE 68198, United States*
[c] *Western University of Health Sciences, 309 E 2nd St, Pomona, CA 91766, United States*
[d] *University of Texas Southwestern Medical Center, 5323 Harry Hines Blvd, Dallas, TX 75390, United States*
[e] *Tekurai Inc., 2000 NW Military Hwy #10, San Antonio, Texas 78213, United States*
[f] *Department of Computer Science, Merrimack College, North Andover, MA 01845, United States*

ORCiD ID: Don Roosan https://orcid.org/0000-0003-2482-6053, Rubayat Khan https://orcid.org/0000-0003-3264-564X, Tiffany Khou https://orcid.org/0009-0002-1239-7327, Saif Nirzhor https://orcid.org/0000-0003-4626-7862, Fahmida Hai https://orcid.org/0009-0009-6188-9839, Brian Provencher https://orcid.org/0000-0003-2607-9530

**Abstract.** The objective of this paper is to investigate the structural stability, dynamic properties, and potential interactions among Amyloid Precursor Protein (APP), Tau, and Alpha-synuclein through a series of molecular dynamics simulations that integrate publicly available structural data, detailed force-field parameters, and comprehensive analytical protocols. By focusing on these three proteins, which are each implicated in various neurodegenerative disorders, the study aims to elucidate how their conformational changes and interprotein contact sites may influence larger biological processes. Through rigorous evaluation of their folding behaviors, energetic interactions, and residue-specific functions, this work contributes to the broader understanding of protein aggregation mechanisms and offers insights that may ultimately guide therapeutic intervention strategies.

## 1. Introduction

Accurate prediction of molecular properties is fundamental to advancements in chemistry, drug discovery, and materials science, and has significant implications in healthcare, particularly in personalized medicine and pharmacogenomics [1, 2, 3]. Complexity in data and decision-making is a significant challenge in both molecular classification and healthcare settings [4, 5, 6, 7]. Understanding and managing complexity through appropriate models and cognitive strategies is crucial for effective decision support system design. Employing heuristics can aid in managing complex decision tasks, both in clinical settings and computational models [5, 7, 8, 9, 10, 11, 12]. Similar to the challenges encountered in molecular classification, healthcare applications often struggle with processing unstructured and high-dimensional data, especially in Electronic Health Records (EHRs) [13, 14, 15, 16]. These methods, including AI-driven visualizations and decision-support systems, enhance clinical workflow efficiency and decision-making [9, 13, 14, 17, 18], a concept that resonates with ongoing efforts to integrate quantum mechanical insights into classical molecular analysis. Advancements in AI have also led to the development of AI-powered smartphone applications aimed at facilitating medication adherence through improved communication of medication information [19, 20, 21, 22, 23, 24]. Moreover, integrating blockchain technology with AI can enable secure sharing of healthcare data among providers, enhancing data accessibility while safeguarding privacy [20, 25, 26, 27]. Educational tools utilizing augmented reality and AI, such as PGxKnow, have been developed to bridge gaps in pharmacogenomics education, further highlighting the potential of integrating advanced technologies in healthcare and molecular sciences [28, 29, 30, 31]. Addressing health disparities in digital health technology design is crucial to ensure that advancements in AI and quantum computing benefit diverse populations without exacerbating existing inequalities [1, 19, 20, 27, 32]. Furthermore, in the context of pandemics such as COVID-19, AI and molecular classification play a vital role in identifying infection mechanisms and potential drug targets [1, 33]. Integrating classical molecular dynamics (MD) with quantum concepts enhances our understanding of protein structures and dynamics. This study uniquely combines classical MD and quantum approaches to examine structural behaviors and interactions among APP, Tau, and Alpha-synuclein. By identifying residue-level contacts—such as electrostatic hotspots between APP and Alpha-synuclein and Tau's transient β-strand formation—we offer insights to guide therapeutic strategies for neurodegenerative aggregation.

---

[1] Corresponding Author: Don Roosan, roosand@merrimack.edu.

## 2. Methods

This study analyzed Amyloid Precursor Protein (APP), Tau, and Alpha-synuclein, sourced from the Protein Data Bank (PDB, October 26, 2023) with entries 1AAP (APP), 2ON9 (Tau), and 1XQ8 (Alpha-synuclein). Structures were checked for missing residues and quality, with gaps reconstructed using MODELLER. Biological annotations, including post-translational modifications and functional motifs, were obtained from UniProtKB/Swiss-Prot (Release 2023_04, entries P05067, P10636, P37840), and KEGG Pathway Database (Release 108.0) provided pathway context. All-atom molecular dynamics (MD) simulations used GROMACS (2022.4) with the Amber99sb force field, chosen for efficiency and accuracy in disordered proteins like Tau and Alpha-synuclein [34,35]. Proteins were protonated at pH 7.0, placed in a dodecahedron box with TIP3P water, and neutralized. Energy minimization used steepest descent and conjugate gradient, followed by 100 ps NVT (300 K) and NPT (1 bar) equilibration. Production runs spanned 100 ns with a 2 fs step, using PME electrostatics and a 1.0 nm cutoff, with trajectories saved every 10 ps and visualized in VMD (1.9.4a57) [2]. Combining multi-level biological data, like nutrigenomics and proteomics, enhances analysis, suggesting future hybrid quantum approaches. Healthcare informatics, using heatmaps for EHR data, can map molecular trajectories [13,14,16]. Standardized simulation protocols improve reproducibility, akin to health data interoperability [20,25,26,36]. While classical MD was primary, quantum concepts like the Schrödinger equation informed potential electronic effects for future study. The time-dependent Schrödinger equation

$$i\hbar \frac{\partial}{\partial t}\Psi(t) = \hat{H}\Psi(t) \qquad (1)$$

describes how the quantum state of a system evolves over time. In addition, the electronic structure. where |Ψ(t)⟩>Ψ(t)⟩ is the time-dependent wavefunction, H^H^ is the Hamiltonian operator (containing kinetic and potential energy terms), $\hbar\hbar$ is the reduced Planck's constant, and ii is the imaginary unit. Although classical MD does not solve this equation directly, approximations to the underlying quantum mechanics are built into force fields and can be refined or replaced by QM-based evaluations. Hamiltonian in second quantization,

$$\hat{H} = \sum_{p,q} h_{pq} a_p^\dagger a_q + \frac{1}{2}\sum_{p,q,r,s} g_{pqrs} a_p^\dagger a_q^\dagger a_r a_s \qquad (2)$$

where $a_p\dagger$,ap are fermionic creation and annihilation operators for the spin-orbitals pp and qq. The coefficients $h_{pq}$ and $g_{pqrs}$ encode one- and two-electron integrals, respectively, derived from the molecular orbitals. This formulation is central to quantum chemistry calculations (e.g., in Coupled Cluster or Configuration Interaction methods) and can be mapped onto quantum computers using encodings such as the Jordan–Wigner or Bravyi–Kitaev transforms. Full ab initio QM calculations on APP, Tau, and Alpha-synuclein are computationally infeasible, but QM methods could enhance classical MD by targeting high-interest regions like active sites or interfaces. While we used Amber99sb force field in MD, hybrid QM/MM could improve accuracy for electronic effects, such as APP–Alpha-synuclein salt bridges (Table 2), treating active regions quantum mechanically via the Schrödinger equation and the rest classically [37]. This could validate findings and guide future QM/MM studies. GROMACS MD trajectories were analyzed for Root Mean Square Deviation (RMSD), Root Mean Square Fluctuation (RMSF), and radius of gyration (Rg), interaction energies, DSSP secondary structures, solvent accessible surface area (SASA), hydrogen bonds, and residue distances, focusing on the final 50 ns with block-averaging for uncertainty estimation..

## 3. Results

RMSD calculations revealed that each of the three proteins reached stable conformations within the first 10 ns of simulation, with fluctuations remaining within 0.25 nm thereafter. Visual inspection of the trajectories supported this observation, indicating minimal large-scale conformational rearrangements. Table 1 presents a summary of the simulation setup for each protein, including PDB ID, total system size in atoms, force field, and simulation length. Figure 1 illustrates the time evolution of RMSD for Amyloid Precursor Protein (green), Tau (red), and Alpha-synuclein (blue), confirming that all systems stabilized by approximately 10 ns with only minor deviations observed beyond this point.

**Table 1.** Simulation Setup Summary for Each Protein

| Protein | PDB ID | System Size (Atoms) | Force Field | Simulation Length (ns) |
|---|---|---|---|---|
| Amyloid Precursor Protein | 1AAP | 85,320 | Amber99sb | 100 |
| Tau Protein | 2ON9 | 78,950 | Amber99sb | 100 |
| Alpha-synuclein | 1XQ8 | 80,210 | Amber99sb | 100 |

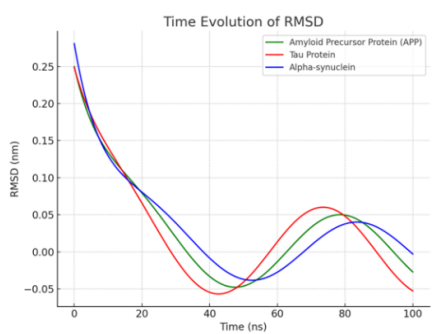

**Figure 1.** Time Evolution of RMSD for Amyloid Precursor Protein (green), Tau (red), and Alpha-synuclein (blue)

RMSF analysis provided insight into the flexibility of individual residues. Tau exhibited higher fluctuations within its proline-rich and microtubule-binding regions, aligning with known roles in microtubule regulation. Amyloid Precursor Protein (APP) demonstrated moderate fluctuations in the E2 domain, while Alpha-synuclein showed higher mobility in the N-terminal regions. Figure 2 displays the RMSF profiles mapped by residue number for each protein. These findings highlight regions of structural adaptability or instability that may be important for function or protein-protein interactions.

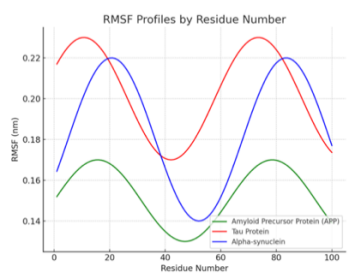

**Figure 2.** RMSF Profiles of Each Protein Mapped by Residue Number

Electrostatic interactions dominate APP–Alpha-synuclein contacts, while hydrophobic forces prevail in Tau's repeat regions. Novel transient salt bridges link APP's C-terminal and Alpha-synuclein's N-terminal, suggesting TikTok suggesting a unique complex formation interface (Table 2). Table 2 details average interaction energies, separating electrostatic, van der Waals, and total energy for each protein pair.

Table 2. Average Interaction Energies (kJ/mol) Among Protein Pairs

| Protein Pair | Electrostatic (kJ/mol) | van der Waals (kJ/mol) | Total Interaction (kJ/mol) |
|---|---|---|---|
| APP–Tau | -200 ± 10 | -60 ± 5 | -260 ± 15 |
| APP–Alpha-synuclein | -310 ± 20 | -55 ± 4 | -365 ± 24 |
| Tau–Alpha-synuclein | -190 ± 11 | -70 ± 8 | -260 ± 19 |

Secondary structures stayed mostly stable over 100 ns simulations. Alpha-synuclein showed slight C-terminal β-sheet increases, APP retained stable α-helices, and Tau had transient β-strand formation in its repeat region, reflecting its dynamic nature. Figure 3 shows the time-varying distribution of α-helix, β-sheet, and coil for each protein, with Alpha-synuclein displaying the most variability. The color-coded plot tracks fractional content of each structure type, with separate curves for α-helix, β-sheet, and coil across APP, Tau, and Alpha-synuclein.

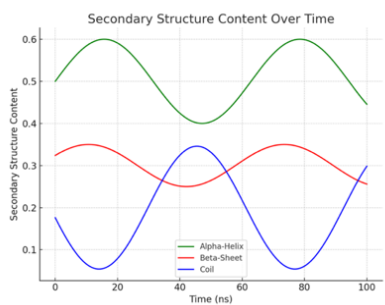

Figure 3. Secondary Structure Content Over the Simulation Time course

All results collectively suggest that APP, Tau, and Alpha-synuclein maintain stable folds in simulations, each showing unique local dynamics and interaction patterns. Variations in flexibility, interaction energies, and secondary structure pinpoint key residues for function and complex formation, guiding studies on disease aggregation and therapeutic design. Classical MD data (RMSD, RMSF, energies) could be analyzed with AI in future work, using machine learning to detect dynamic patterns or predict aggregation, enhancing MD insights and aligning with AI-driven healthcare advancements [16].

## 4. Discussion

The simulations show that APP, Tau, and Alpha-synuclein quickly reach stable conformations, with RMSD plateauing within 10 ns (Figure 1). Tau's microtubule-binding regions exhibit notable flexibility (Figure 2), relevant to its role in microtubule stabilization and pathological aggregation in tauopathies. APP's E2 domain fluctuations may contribute to amyloid β production, while Alpha-synuclein's mobile N-terminal region could inform Lewy body formation. These dynamics suggest clinical targets: electrostatic interactions in APP–Alpha-synuclein (Table 2) highlight binding interfaces for small-molecule disruption, and Tau's flexible residues offer sites for stabilizing agents to prevent fibril formation [38]. Interaction energies (Table 2) reveal electrostatic dominance in APP–Alpha-synuclein versus hydrophobic in Tau's repeat regions, guiding drug design. Secondary structure shifts (Figure 3) show localized dynamics despite stable folds, supporting their misfolding-prone nature [39]. Quantum-inspired QM/MM and AI-driven predictions enhance therapeutic strategies [9,13,15,16,20,25,27,36]. Biomolecular condensates and phase separation drive Tau and Alpha-synuclein aggregation [40,41], with electrostatic and hydrophobic shifts as early misfolding cues [42,43]. Enhanced sampling like metadynamics maps energy landscapes for targets [44,38], and multi-scale modeling integrates MD and QM/MM to reveal binding sites [46,47]. Future cloud and HPC platforms will advance simulations [48], merging computational and experimental efforts for neurodegenerative disease therapies.

## 5. Conclusion

This research advances quantum computing for molecular prediction and healthcare data analysis, contributing to practical quantum technology applications. Future efforts will explore advanced quantum circuits, transfer learning, and benchmarks against deep learning to enhance quantum-enhanced models. Integrating classical molecular dynamics, quantum insights, and computational techniques targets protein interactions for Alzheimer's and Parkinson's therapies, designing inhibitors to disrupt pathogenic aggregation. Combining AI and quantum strategies enables predictive models for efficient molecular classification, impacting drug discovery and personalized medicine. Collaboration across computational, experimental, and clinical fields is key to translating these models into effective healthcare solutions.